\documentclass{elsart}
\usepackage{graphicx,amssymb}
\renewcommand{\today}{}
\begin{document}

\begin{frontmatter}

\title{Demonstration of fundamental mode only propagation in highly
multimode fibre for high power EDFAs}
\author[qinetiq]{C. D. Stacey},
\author[qinetiq]{R. M. Jenkins},
\author[prl]{J. Banerji} and
\author[rh]{A R Davies}
\address[qinetiq]{Optronics Research and Consulting, QinetiQ, St.Andrews Rd., Malvern, WORCS WR14 3PS, U. K.}
\address[prl]{Quantum Optics and Quantum Information Group, Physical Research Laboratory, Navrangpura, Ahmedabad 380 009, India}
\address[rh]{Department of Computer Science, Royal Holloway, University of London, Egham, Surrey TW20 0EX, U. K.}
\begin{abstract} The use of short lengths of large core phosphate
glass fibre, doped with high concentrations of Er or Er:Yb
represents an attractive route to achieving high power erbium doped
fibre amplifiers (EDFAs) and lasers (EDFLs). With the aim of
investigating the potential of achieving diffraction limited output
from such large core fibres, we present experimental results of
fundamental mode propagation through a 20 cm length of passive 300
\ensuremath{\mu}m core multimode fibre when the input is a
well-aligned Gaussian beam. Through careful control of fibre
geometry, input beam parameters and alignment, we measured an output
$M^{2}$ of $1.1 \pm 0.05$. The fibre had a numerical aperture of
0.389, implying a V number of 236.8. To our knowledge, this is the
largest core fibre through which diffraction limited fundamental
mode propagation has been demonstrated. Although the results
presented here relate to undoped fibre, they do provide the
practical basis for a new generation of EDFAs and EDFLs.
\end{abstract}
\today
\end{frontmatter}

\section{Introduction}
Conventional EDFAs based on single mode fibre have limited power
handling characteristics. For many industrial and military
applications these power levels are prohibitively low. By using
larger core multimode fibre, power levels can be scaled-up,
avoiding non-linear effects (stimulated Brillouin and Raman
scattering) and catastrophic optical damage in the core and at the
end facets. Although many research groups have investigated
amplifiers based on multimode fibres \cite{1,2}, most of the
fibres investigated have had core diameters less than 50
\ensuremath{\mu}m; and more often less than 30 \ensuremath{\mu}m.
In practice, bend-loss or tapered fibre sections are used to
filter out higher order modes. Koplow \cite{1} showed that the
bend-loss technique becomes less effective for larger core fibres.
Higher order mode suppression in a 100 \ensuremath{\mu}m core
fibre is at least a factor of 10 lower than that observed in the
more commonly used 25-50 \ensuremath{\mu}m core fibres.With regard
to the use of tapers, these not only filter out higher order modes
but also power. This limits the amplifier gain as shown by Minelly
\cite{3}.

In earlier work we have demonstrated high fidelity fundamental mode
propagation through hollow multimode waveguides with diameters of
thousands of microns \cite{4,5}. This was achieved by suitable
control of the launch conditions and waveguide linearity. Recently,
we have proposed the extension of this concept to the realisation of
very large mode area (VLMA) EDFAs with core diameters of 300
\ensuremath{\mu}m or more. Similar work has been carried out by W.
S. Wong et al \cite{5.1} who recently demonstrated this propagation
in VLMA fibre using photonic crystal fibre. This exploited the
natural mode selection which arises through the use of large air
holes in the fibre cladding. Having an effective mode field diameter
of 42.5 \ensuremath{\mu}m, the fundamental mode was excited and
propagated over an impressive 4 m fibre length.

In 2004, Qiu {\it et al} \cite{6} demonstrated 4 W output power, in
single mode, from a 7 cm length of Er:Yb co-doped phosphate glass
fibre. The same research group (and others) has shown that phosphate
glass fibre is much more soluble to erbium and ytterbium ions and
thus provides significantly higher pump absorption per unit length
of fibre \cite{7}. The result is that significant gain and output
powers can be attained from comparatively short lengths of fibre. It
is this approach, in conjunction with the aforementioned single mode
propagation in multimode waveguides that is of particular interest.
In this letter we describe the potential of VLMA EDFAs and present
experimental measurements of the passive fundamental mode
propagation characteristics of a 300 \ensuremath{\mu}m core
multimode fibre using a well-aligned Gaussian as the input beam.
\section{Background}

The V-number of a fibre is an indicative measure of how many higher
order modes can propagate through the fibre \cite{8}. It is given by
\begin{equation}\label{eq1} V=\frac{2\pi }{\lambda} a\times (\rm
NA)\end{equation}

where $\lambda$ is the wavelength of the radiation, $a$ is the
fibre core radius and NA is the numerical aperture relating to the
refractive index step between the core and the cladding. To force
a fibre to only support the propagation of the lowest order
fundamental mode, the V-number must be kept at, or below 2.405
\cite{8}. If the core diameter is increased whilst maintaining the
same index-step, a higher V-number results and the fibre becomes
multimode. With respect to maintaining single mode
characteristics, an increased core diameter can be offset, up to a
point, by a reduced refractive index step as demonstrated by
Taverner \cite{9} for a 20 \ensuremath{\mu}m core fibre. However,
this is only effective in fibres up to a certain core size and
often requires additional fibre bending to achieve single mode
output. To fabricate a single mode fibre with a core diameter of
100 \ensuremath{\mu}m using this approach is currently beyond
fibre processing constraints.

Even though a fibre might have a multimode nature, it is still
possible to excite just the fundamental mode. This depends on the
input beam profile and the way the beam is injected into the fibre.
Since our primary objective is to demonstrate the propagation of the
fundamental mode and not any particular input beam, we have chosen a
Gaussian beam as the input field. In this case, fundamental mode
propagation can be achieved by holding the fibre rigid and straight,
and controlling the alignment with respect to the injected Gaussian
beam. In this manner, fundamental mode propagation can be maintained
over some distance. This assumes that there are no perturbations
(periodic or non-periodic) in the fibre and that the input
beam-width is carefully selected to ensure that the value of the
overlap integral between the single mode input field and the
fundamental mode of the fibre is maximised. Figure 1 shows a plot of
fundamental mode power coupling coefficient $C(\gamma)$ as a
function of $\gamma=w/a$, the ratio of input beam (assumed
Gaussian)waist $w$ and fibre core radius $a$, for a 300
\ensuremath{\mu}m core diameter. Also shown are the values of
$C(\gamma)$ for various values of $\gamma$ near optimum coupling. It
is evident that a Gaussian input having a beam waist corresponding
to $\gamma\sim 0.65$ is required for optimum coupling of power into
the fundamental mode of the fibre. It can be shown that a more
precise value of $\gamma$ for this purpose is given by
$\gamma=0.6463$. In our experiment, the value of $\gamma$ was chosen
to be 0.65.

\begin{figure}[htbp]
\begin{center}
\includegraphics*[width=13.0cm]{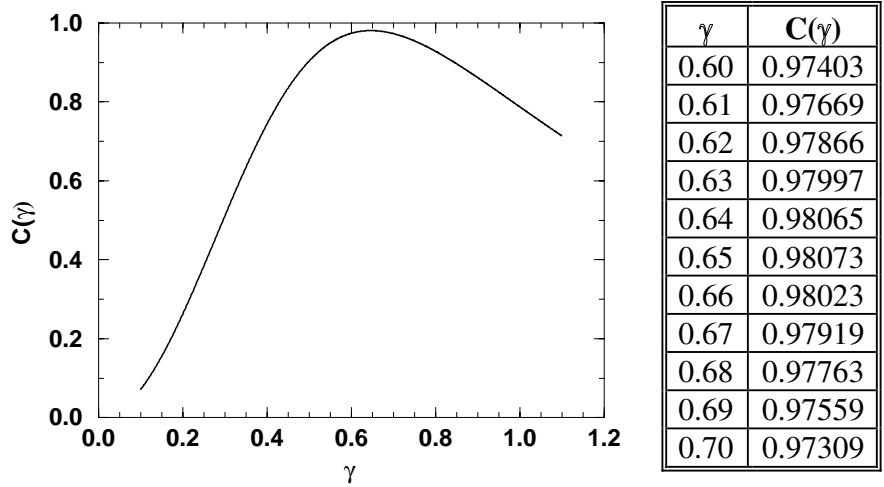}
\end{center}
  \caption{Power coupling coefficient from free space to the fundamental mode of the fibre, as a function of input field beam waist.}
  \label{f1}
  \end{figure}

Assuming an input beam from a laser source has a Gaussian form, it
can be shown \cite{10} that the magnitude of the field overlap
integral with the sum of all the guided modes of the fibre (having
forms of Bessel functions) has a value of 0.99. This implies that
98\% (absolute square of overlap integral) of input power is coupled
to guided modes of the fibre. If the fibre is aligned to the input
in both angular and longitudinal senses, then it can be further
shown through modal analysis that 99.8 \% of the light coupled to
the guided modes of the fibre is, in fact, coupled to the
fundamental mode -- the remaining 0.2 \% to low order modes
LP$_{02}$ and LP$_{03}$. To illustrate mode-coupling as a function
of misalignment further, figure 2 shows the power coupling
coefficient calculated as a function of both angular and
translational misalignment of fibre with respect to input field.
These results are numerical predictions based on a 300
\ensuremath{\mu}m core fibre with NA of 0.389. The coupling
coefficient is defined as the modulus squared of the overlap
integral between a Gaussian input field and each of the indicated
modes at the misalignments shown.
\begin{figure}[htbp]
\begin{center}
\includegraphics*[width=13.0cm]{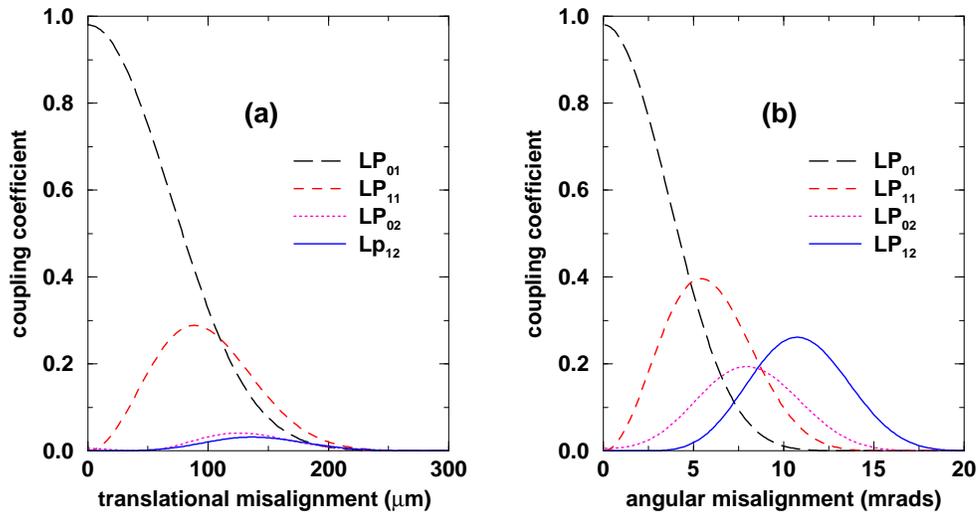}
\end{center}
\caption{Power coupling coefficients to each of the four lowest
order modes, numerically calculated for a 300 \ensuremath{\mu}m core
fibre of NA=0.389 (a) translational misalignment (b) angular
misalignment.}\label{f2}
\end{figure}
It is evident that the tolerances on alignment are very high for
such a multimode fibre, with an angular tolerance of greater than 3
\ensuremath{\mu}m required to suppress excitation of LP11. Figure
2(a) shows that coupling to all guided modes diminishes with
increasing misalignment, with very little coupling to any mode
beyond 200  \ensuremath{\mu}m. In contrast, figure 2(b) shows
significant higher order mode excitation for angular misalignments
up to and exceeding 20 mrads. Although not shown, significant
higher-order mode coupling (coupling coefficient \ensuremath{\sim}
 0.2) occurs at angles exceeding 100 mrad. It is therefore of vital
importance that controls are taken over these sources of
misalignment and mode mismatch. We end this section by noting that
the above analysis is valid only for a Gaussian input beam. The next
section will cover the experimental realization of the fundamental
mode VLMA fibre principle.
\section{Experimental configuration and results} As a first step in
assessing the concept, we set out to investigate if it was possible
to achieve high fidelity fundamental mode excitation and propagation
through a passive multimode fibre. In order to achieve and maintain
high fidelity fundamental mode propagation in a multimode fibre, the
launch conditions and the linearity of the fibre must be
appropriately controlled. The fibre needs to be kept as straight as
possible, as bending will cause the fundamental mode field to be
coupled to higher order modes. Fibre-guides with a ''V'' shaped
cross-section were milled into the surface of a machinable ceramic
(Macor) substrate using computer-aided machining techniques
\cite{4}. The fibre-guides ranged in length from 5 cm to 20 cm. Lids
of the same material held the fibres in position. The fibre-guides
accommodate fibres with cladding diameters ranging from 125
\ensuremath{\mu}m up to 1500 \ensuremath{\mu}m. A fibre (mounted in
such a fibre-guide) was held in position on a commercially available
Newport translation stage (565-XYZ-TILT) with five degrees of
freedom: three translational (with a resolution of \ensuremath{\sim}
0.1 \ensuremath{\mu}m), and two angular (with a resolution of
\ensuremath{\sim} 2 \ensuremath{\mu}rad). The fibre used in this
experiment was supplied by Thorlabs (FT-300-EMT) and had a 300
\ensuremath{\mu}m core diameter and 325 \ensuremath{\mu}m cladding
diameter. The value of NA was 0.389. The tolerance on the outer
diameter of the polymer coating was 30 \ensuremath{\mu}m, which was
a factor of three worse than that of the silica inner cladding. For
this reason, the polymer coating was removed in order to minimise
micro bending from this source.

Prior to mounting in the fibre-guide, the fibre end faces were
cleaved and polished to a surface flatness of \ensuremath{\sim}
\ensuremath{\lambda}/30 at 1550 nm. This prevented phase
distortion of the fundamental mode input field.

The large V number of the fibre leads to a strong confinement of the
field within the core. In this context, the input beam was chosen to
have a diameter which ensured optimum coupling to the fibre. Through
numerical analysis of overlap integrals between input field and the
fundamental mode of the fibre, it was predicted that an input beam
with a 1/e$^{2}$ mode field diameter of 0.646 of the fibre core
diameter, would provide optimum power coupling to the fundamental
mode of the fibre.

A fibre-coupled 1550 nm diode laser (Alcatel A1905 LMI 30mW) was
focussed via a 10X microscope objective to achieve a beam waist of
195 \ensuremath{\mu}m (0.65 X 300 \ensuremath{\mu}m). At the output
end of the fibre, a second microscope objective was used to produce
a magnified image of the emerging field onto a ``Gentec BeamR'' beam
profiler. A Hamamatsu vidicon camera was also used for additional
analysis of the fibre output. A schematic of the experimental
configuration is illustrated in figure 3. The output from the fibre
laser obviously couples to some free-space Hermite-Gaussian mode
which we assume to very closely approximate to a quasi-Gaussian
field profile at the multimode fibre input.

\begin{figure}[htbp]
\begin{center}
\includegraphics*[width=13.0cm]{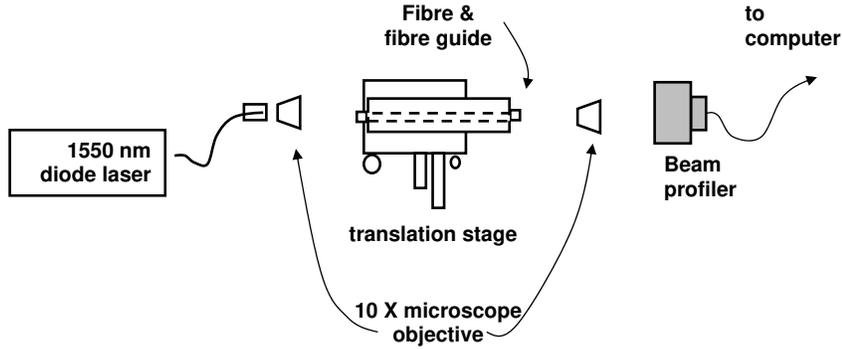}
\end{center}
\caption{Experimental setup of fibre, coupling optics and beam
profiler.}\label{f3}
\end{figure}

Alignment of the fibre was by no means trivial and to obtain the
desired output required extensive fine-tuning of translational and
angular positions. Figure 4 shows the beam profiler plots of the
field output of a well-aligned 20 cm length of 300 \ensuremath{\mu}m
core fibre.
\begin{figure}[htbp]
\begin{center}
\includegraphics*[width=13.0cm]{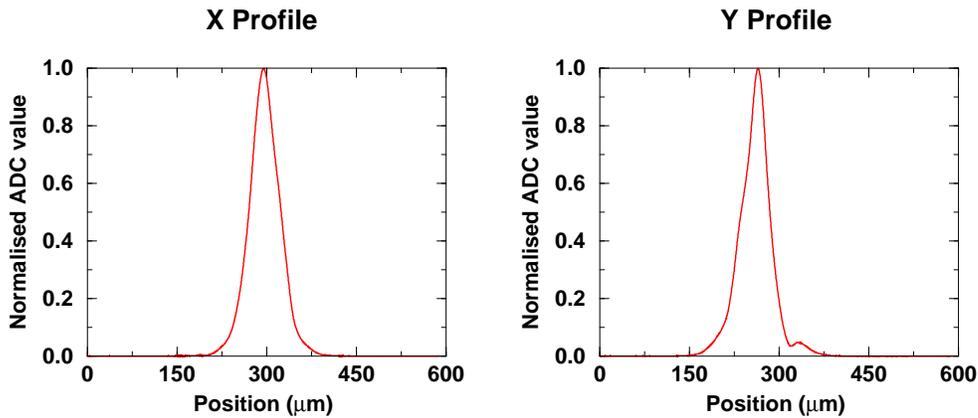}
\end{center}
\caption{Measured beam waist profiles in orthogonal orientations X
and Y.}\label{f4}
\end{figure}
As illustrated, the output beam
profile had a near Gaussian-form. A measured M$^{2}$ value of $1.1
\pm 0.05$ confirmed that efficient fundamental mode propagation
had been achieved in practice. The profiles shown represent
orthogonal axes of measurement. The slight asymmetry between them
is most likely due to some higher order mode content, which is of
such low amplitude that it does not have significantly adverse
effect on the measured M$^{2}$ parameter. Similar values of
M$^{2}$ were measured for fibre lengths of 5, 10, and 15 cm. These
experiments were repeated for a 105 \ensuremath{\mu}m core
diameter multimode fibre (again, commercially available from
Thorlabs -- AFS105/125Y) which gave comparable results, i.e.
M$^{2}$ of \ensuremath{\sim}1.1 for lengths up to 20 cm.

These results are very encouraging, offering a proof-of-principle
that highly multimode fibre, capable of supporting in excess of
28,000 (\ensuremath{\sim}V$^{2}$/2) modes can be configured in
conjunction with a suitable input field, such that only its
fundamental mode is excited and propagated throughout its length. We
did not observe any deterioration in beam quality with length for
fibres ranging from 5 cm to 20 cm, indicating longer lengths of
fibre would also support the fundamental mode only. One might well
ask whether the comparatively short fibre lengths considered here
are transmitting the light in the form of a guided mode, or whether
the fibre is so large as to be simply acting as a piece of bulk
silica transmitting the light. However, it is worth noting that the
Rayleigh range of a 195 \ensuremath{\mu}m diameter input beam
entering the fibre is only 1.9 cm. In view of this, we can be
confident that the light is indeed transmitted in a guided mode.

\begin{figure}[htbp]
\begin{center}
\includegraphics*
[width=13.0cm]{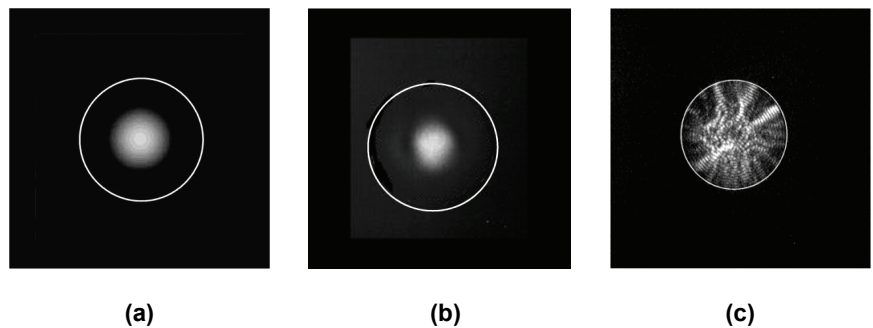}
\end{center}
\caption{(a) Theoretical plot of fundamental mode of 300
\ensuremath{\mu}m core fibre with NA=0.389, (b) fundamental mode
output from 300 \ensuremath{\mu}m core fibre, held straight, well
aligned and polished, (c) Near field fibre output profile from the
same fibre, but kept unpolished and misaligned demonstrating
multimode nature. The white ring on each plot represents the
core/cladding interface.}\label{f5}
\end{figure}

Figure 5 (a) shows an analytically generated image of the field
corresponding to the fundamental mode of a fibre of the same
specification as used here. Figure 5 (b) shows a vidicon camera
image of the near field of the fibre output. This is the same output
as illustrated in figure 4. The white circle encompassing the field
distributions in all plots represents the core cladding interface. A
20 cm length of the same 300 \ensuremath{\mu}m core fibre was
located in a fibre guide of inferior quality, such that the fibre
was not held straight. Additionally, the end faces were cleaved but
not polished. All other parameters were kept the same. The resulting
output field is shown in figure 5 (c), which illustrates the highly
multimode nature of the fibre, with significant higher order mode
content. This also confirms the necessity of the controls we have
taken over fibre linearity and minimisation of mode coupling.

\section{Conclusions}
We have experimentally demonstrated that a 300 \ensuremath{\mu}m
core multimode fibre can support high fidelity fundamental mode
(LP$_{01}$) propagation over lengths of 20 cm or more. Since there
was no variance in the output mode quality with length from 5 cm
to 20 cm, there is every reason to believe that fibre lengths can
be extended to 40 cm and still produce near diffraction-limited
output beams. Although the technical discussions outlined here
relate to amplifier applications, laser configurations should also
be feasible with sufficient controls over preferential mode lasing
\cite{11}.

With regard to non-linear scattering, a 300 \ensuremath{\mu}m
core, 30 cm long EDFA would be capable of handling 30 kW optical
power before reaching threshold for stimulated Brillouin
scattering (SBS). From a thermal damage perspective, it is
accepted that a conservative damage threshold for rare earth doped
fibre is \ensuremath{\sim}1 W/\ensuremath{\mu}m$^{2}$ \cite{12}.
For a 300 \ensuremath{\mu}m core EDFA, this equates to a damage
threshold of almost 70 kW. Although these figures are unlikely to
be practically realised in EDFAs of this nature, they do
illustrate the utility of larger core fibres in overcoming the
limiting factors in conventional fibre power amplifiers. Heat
generation from the quantum defect in EDFAs and fracture limits
would be the first limit reached in fibres such as those presented
here, with a limit of around 800 W expected for a fibre of this
length based on parameters presented in previous work
\cite{13,14}.

Provided rare-earth doped fibres behave similarly to the fibre
described in this letter in terms of mode propagation, this would
appear to offer a way of developing, in a master oscillator power
amplifier (MOPA) configuration, a high power, compact and robust,
single-mode amplifier, without the need for fibre coiling.

\section{Acknowledgements}

We acknowledge the UK Ministry of Defence in the support of this
program.

\newpage
\section*{List of Figure Captions}

Fig. 1 Power coupling coefficient from free space to the fundamental
mode of the fibre, as a function of input field beam waist.

Fig. 2 Power coupling coefficients to each of the four lowest order
modes, numerically calculated for a 300 \ensuremath{\mu}m core fibre
of NA=0.389 (a) translational misalignment (b) angular misalignment.

Fig. 3 Experimental setup of fibre, coupling optics and beam
profiler.

Fig. 4 Measured beam waist profiles in orthogonal orientations X and
Y.

Fig. 5 (a) Theoretical plot of fundamental mode of 300
\ensuremath{\mu}m core fibre with NA=0.389, (b) fundamental mode
output from 300 \ensuremath{\mu}m core fibre, held straight, well
aligned and polished, (c) Near field fibre output profile from the
same fibre, but kept unpolished and misaligned demonstrating
multimode nature. The white ring on each plot represents the
core/cladding interface.
\end{document}